\documentclass[runningheads]{svmult}

\usepackage{makeidx}   
\usepackage{graphicx}  
\usepackage{subeqnar}  
\usepackage{multicol}  
\usepackage{physprbb}  
\makeindex             

\newcommand{\greeksym}[1]{{\usefont{U}{psy}{m}{n}#1}}
\newcommand{\umu}{\mbox{\greeksym{m}}}

\begin{document}
\title*{Future Perspectives at CERN}
\toctitle{Future Perspectives at CERN}
%
%
\titlerunning{Future Perspectives at CERN}
%
\author{John Ellis\inst{1}}
%
\authorrunning{John Ellis}
%
%
\institute{Theoretical Physics Division, CERN, Geneva, Switzerland}

\maketitle              

\begin{abstract}
Current and future experiments at CERN are reviewed, with emphasis on
those relevant to astrophysics and cosmology. These include experiments
related to nuclear astrophysics, matter-antimatter asymmetry, dark
matter, axions, gravitational waves, cosmic rays, neutrino oscillations,
inflation, neutron stars and the quark-gluon plasma. The centrepiece of
CERN's future programme is the LHC, but some ideas for
perspectives after the LHC are also presented.
\end{abstract}
\begin{center}
{CERN-TH/2002-119 ~~~~~~~~~~ astro-ph/0206054}\\
{\it Talk given at the CERN-ESA-ESO Symposium, M{\"u}nchen, April 2002}\\
\end{center}
\section{Outline}

The scientific mission of CERN is to provide Europe with unique
accelerators for the study of the fundamental particles of matter and the
interactions between them. The scientific programme of CERN for the next
decade is centred on the LHC accelerator, which is scheduled for
completion in 2006, so that its experiments can start taking in 2007. A
description of the LHC scientific programme is the centrepiece of this
talk. The motivations for this and other new accelerators provided by
ideas about possible physics beyond the Standard Model were discussed
earlier at this meeting~\cite{Ellis}.

Between now and the startup of the LHC, CERN has a very limited programme
of running experiments. However, scientific diversity at CERN is enhanced
by a number of recognized experiments, that do not use the CERN
accelerators and are not supported by CERN, but whose scientists are
allowed to use other CERN facilities. In parallel with the construction of
the LHC, CERN is also preparing to send a long-baseline neutrino beam to
the Gran Sasso underground laboratory in Italy, in a special programme
largely supported by extra contributions from interested countries. These
are described before the LHC programme.

In 
the longer term, CERN has started thinking about its future prospects
beyond the LHC. Various options have been proposed, including upgrades of
the LHC, a concept for a multi-TeV electron-positron collider called CLIC,
a neutrino factory, and a possible role in space experiments. These are
mentioned at the end of this talk, with particular emphasis on CLIC.

At each stage in this talk, the relevances of CERN experiments to
astrophysics, cosmology and space science are emphasized. The symbiotic
relationships between particle physics and these subjects, that motivated
this meeting, are amply reflected in the many connections between
microphysics and macrophysics revealed in this brief survey.

\section{Present CERN Experiments}

\begin{figure}[h]
\begin{center}
\includegraphics[width=.5\textwidth]{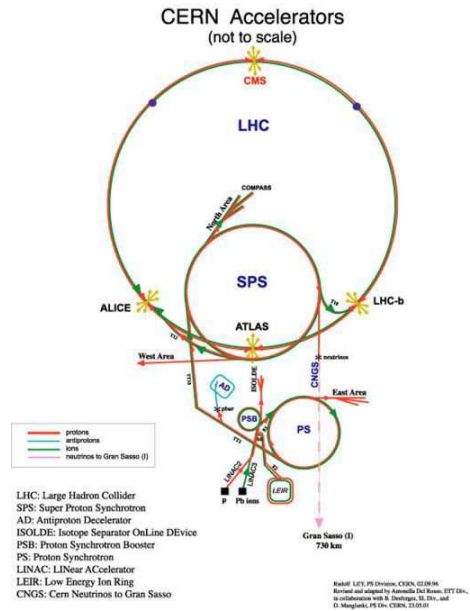}
\end{center}
\caption[]{\it
Map of the CERN accelerators, showing how the LHC will be fed from the 
smaller SPS and PS rings via the transfer lines TI~2, TI~8. Also shown are 
the locations of the ALICE, ATLAS, CMS and LHCb experiments, as well as 
the starting-point of the CNGS beam and the antoproton decelerator (AD).
}
\label{fig:LHC}
\end{figure}

In addition to the LHC accelerator that is currently under construction,
as seen in Fig.~\ref{fig:LHC}, CERN has a number of lower-energy
accelerators operating at energies between a few hundred MeV and several
hundred GeV. Protons at the lowest energies feed the ISOLDE facility,
CERN's source of radio-active ions. Several experiments at ISOLDE address
astrophysical issues, including a search for axions and massive neutrinos,
studies of neutron-rich isotopes relevant to the supernova $r$-process,
etc.~\cite{ISOLDE}.

\begin{figure}[h]
\begin{center}
\includegraphics[width=.7\textwidth]{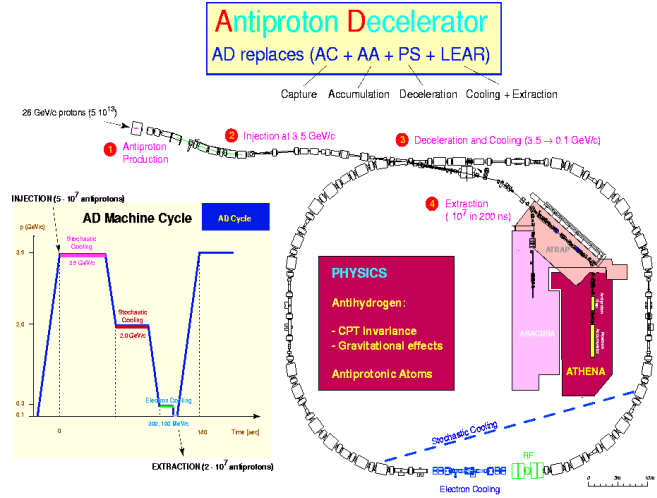}
\end{center}  
\caption[]{\it
Layout of the Antiproton Decelerator (AD) complex, including the three
experiments ASACUSA, ATHENA and ATRAP.
}
\label{fig:AD}
\end{figure}

Protons from CERN's next-lowest-energy accelerator, the PS, are used
partly to make antiprotons. These are in turn slowed down in CERN's
antiproton decelerator (AD) and used to manufacture antihydrogen
atoms~\cite{Hbar} in the world's first antimatter factory~\cite{AD},
illustrated in Fig.~\ref{fig:AD}.  Their numbers are insufficient by many
orders of magnitude to drive spaceships \`a la Star Trek, but they can be
used to test matter-antimatter asymmetry in the form of CPT violation,
with unprecedented accuracy. Scenarios for generating the baryon asymmetry
of the Universe usually rely on the breaking of matter-antimatter symmetry
via CP violation, but it has sometimes been suggested that the violation
of CPT might also play a role. The AD and its associated experiments will
provide some pointers on such ideas, by using laser spectroscopy to probe
for differences in the energy levels of hydrogen and antihydrogen atoms.

CERN's current highest-energy accelerator, the SPS, is mainly used to test
components for the LHC experiments, but also has a limited research
programme. For example, the COMPASS experiment~\cite{COMPASS} is
contributing to the understanding of the proton spin, which is relevant to
calculations of the scattering of cold dark matter particles. The NA48
experiment~\cite{NA48} has been studying CP violation in kaon decays,
establishing its presence directly in decay amplitudes~\cite{epsprime}, as
postulated in many scenarios for baryogenesis. This an important proof of
principle, but baryogenesis would require analogous direct CP violation
the decays of different particles.

\begin{figure}[h]
\begin{center}
\includegraphics[width=.7\textwidth]{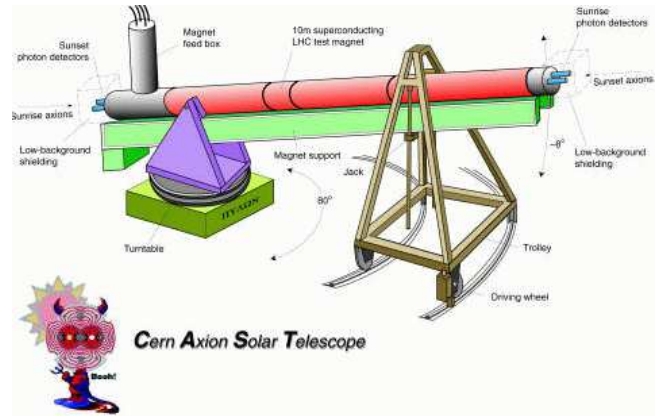}
\end{center}  
\caption[]{\it
Design of the CAST experiment, which will use a surplus LHC magnet to search
for solar axions.
}
\label{fig:CAST}
\end{figure}

CERN is also conducting one non-accelerator experiment of interest to
astrophysics, namely the CAST experiment~\cite{CAST} that is searching for
axions from the Sun. Shown in Fig.~\ref{fig:CAST}, it uses a surplus
superconducting LHC magnet to look for axion-to-photon conversion in a
strong magnetic field, and hopes to achieve a sensitivity beyond indirect
astrophysical limits on axions~\cite{Raffelt}.

\section{Experiments Recognized at CERN}

As has already been mentioned, recognized experiments do not use CERN
accelerators, and receive no financial support from CERN, but are allowed
officially to use other CERN facilities, such as office space and the
computer network. These concessions were originally requested by
physicists who were splitting their time between some experiments that use
CERN accelerators and others that do not. CERN accepts the principle of
such time-sharing, and is grateful for the scientific diversity that it
provides, particularly during the pre-LHC period when there are relatively
few new CERN data.

\subsection{Gravitational Waves}

EXPLORER~\cite{EXPLORER} was the first experiment to be recognized at
CERN. It uses CERN's cryogenic facilities to keep cold its large bar
detector for gravitational waves. More recently, CERN has recognized the
LISA experiment~\cite{LISA}, a trio of spacecraft that is now being designed 
to look for lower-frequency, longer-wavelength gravitational waves in space.

\subsection{Astrophysical Antimatter}

\begin{figure}[h]
\begin{center}
\includegraphics[width=.5\textwidth]{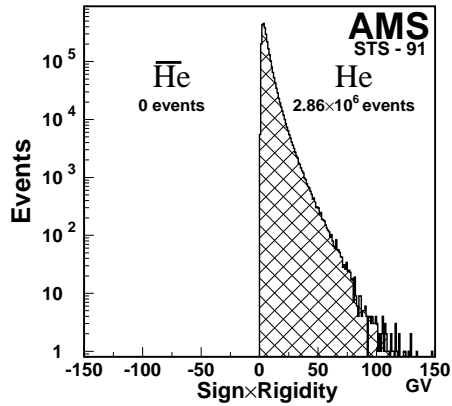}
\end{center}  
\caption[]{\it
The AMS experiment looked for antihelium nuclei during its first flight
on the space shuttle. It found many conventional helium nuclei, but no
antihelium.
}
\label{fig:AMS}
\end{figure}

AMS~\cite{AMS} was the first particle spectrometer to have been sent into
space, on the space shuttle in 1998, and an improved configuration is
currently being prepared to fly again for several years on the
International Space Station. AMS looks, in particular, for antimatter
particles such as positrons, antiprotons and antinuclei, that might be
signatures of antimatter in the Universe, or of dark matter annihilations.
Its first flight made interesting measurements of the Earth's cosmic-ray
albedo~\cite{albedo}, and set a new upper limit on anti-helium in the
primary cosmic-ray flux~\cite{Hebar} (see Fig.~\ref{fig:AMS}) that
constitutes further evidence against a matter-antimatter symmetric
cosmology. Most of the financial support for the AMS detector comes from
European funding agencies, particularly in France, Germany, Italy and
Switzerland, so it is natural to have a centrally-located ground base in
Europe. Two other recognized space experiments that will also look for
astrophysical antimatter are CAPRICE~\cite{EXPLORER} and
PAMELA~\cite{PAMELA}. GLAST~\cite{GLAST} is a satellite gamma-ray
spectrometer able to look, for example, at gamma-ray bursters and/or
energetic photons from the core of our galaxy that might be generated by
the annihilations of dark matter particles.

\subsection{Neutrino Telescopes}

ANTARES~\cite{ANTARES} and NESTOR~\cite{NESTOR} are two underwater
neutrino telescopes that will be looking for high-energy cosmic neutrinos,
as might come from astrophysical sources or the annihilations of dark
matter particles in the Sun and the core of the Earth.

\subsection{Cosmic Rays}

L3+C~\cite{L3+C} is an extension of the L3 experiment, that made
measurements with CERN's LEP accelerator, using its muon detectors and
additional counters to study cosmic rays. L3+C has provided some
interesting measurements of the cosmic-ray muon flux, that help constrain
calculations of the atmospheric neutrino flux, and hence refine the
interpretation of neutrino oscillation experiments. AUGER~\cite{AUGER} is
an ultra-high-energy cosmic-ray experiment being constructed in Argentina
by a team from several continents, with a strong European participation.
As discussed here by Watson~\cite{Watson}, if the AGASA
experiment~\cite{AGASA} is correct, AUGER should be able to gather large
numbers of events from beyond the Greisen-Zatsepin-Kuzmin
cutoff~\cite{GZK}, and tell whether they are due to compact astrophysical
sources, or to the decays of very heavy metastable dark matter particles
(cryptons~\cite{BEN}). There is currently some controversy in the energy
calibrations of ultra-high-energy cosmic rays, which accelerator
experiments at CERN could in principle help resolve, e.g., by a PS or SPS
experiment to remeasure the spectrum of fluorescence light from nitrogen
at different frequencies~\cite{Watson}, and/or by constraining models of
high-energy particle showers using data on hadron production at the
LHC.

\section{Neutrino Beam from CERN to the Gran Sasso Laboratory}

CERN is currently constructing a beamline for sending neutrinos the 730~km
to the Gran Sasso underground laboratory in Italy, called the
CNGS project~\cite{CNGS}, whose starting-point is illustrated in 
Fig.~\ref{fig:LHC}.
Experiments on atmospheric neutrinos suggest strongly that $\mu$ neutrinos
oscillate mainly into $\tau$ neutrinos~\cite{SK}, but direct experimental
proof is still lacking. The energy $E_\nu \sim 20$~GeV of the beam to be
sent from CERN to the Gran Sasso is optimized for $\tau$ production via
the charged-current reaction $\nu_\tau + N \to \tau + X$. Civil
engineering was started in 2000, and the beam may be commissioned in
2006~\cite{others}.

The first experiment to be approved for the long-baseline beam was
OPERA~\cite{OPERA}, which will use emulsion techniques with high spatial
resolution to identify events in which $\tau$ leptons are produced. The
ICARUS collaboration~\cite{ICARUS} proposes to build a 3000-tonne liquid
argon calorimeter in the Gran Sasso underground laboratory, having already
demonstrated the feasibility of the technique using a 600-tonne pilot
module, and got approval to install it at Gran Sasso.

In the longer run, other experiments might want to use the CERN-Gran Sasso
beam. For example, it has recently been proposed~\cite{Dydak} to place a
detector in the Gulf of Taranto, just off the beam axis~\cite{Velasco},
where it would be very sensitive to $\nu_\mu \to \nu_e$ transitions.

\section{LHC}

The Large Hadron Collider (LHC), under construction for installation in
the 27~km tunnel previously used for LEP, is primarily designed to deliver
proton-proton collisions with a centre-of-mass energy of 14~TeV with a
luminosity of $10^{34}$~cm$^-2$s$^{-1}$~\cite{LHC}. It will also be able
to collide lead ions at a centre-of-mass energy of 1.2~PeV with a
luminosity of $10^{27}$~cm$^-2$s$^{-1}$.

\begin{figure}[h]
\begin{center}
\includegraphics[width=.5\textwidth]{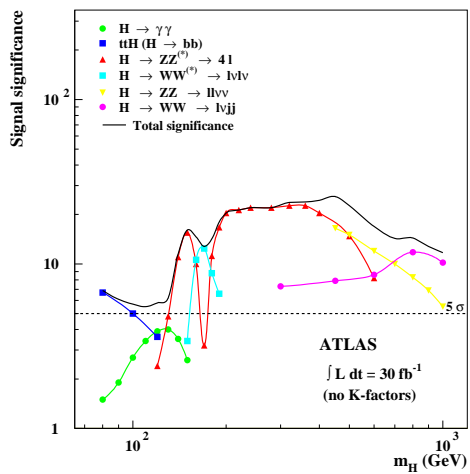}
\end{center}  
\caption[]{\it
Either of the major LHC experiments will be able to discover the Higgs boson
of the Standard Model in a variety of different decay modes, with a total
significance exceeding 10 standard deviations.
}
\label{fig:Higgs}
\end{figure}

The main objective of the proton-proton programme is to explore physics in
a new energy range from 100~GeV to a TeV and beyond. This is the energy
range where the origin of particle masses is expected to be revealed, and
one of the principal preys of the LHC will be the Higgs boson, or whatever
accomplishes its task of giving masses to the elementary particles.  The
prospects of finding the Higgs boson are good, as seen in
Fig.~\ref{fig:Higgs}~\cite{TDRs}. Discovery of such an elementary scalar
boson would also be interesting for cosmologists, since it is the
prototype for the inflaton~\cite{inflation}.

\begin{figure}[h]
\begin{center}
\includegraphics[width=.4\textwidth]{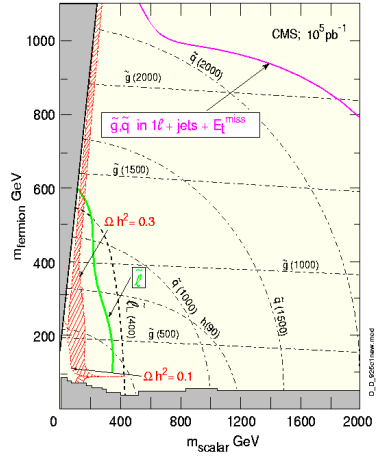}
\end{center}  
\caption[]{\it
Either of the major LHC experiments will be able to discover supersymmetry
with high significance throughout most of the parameter space where the
lightest supersymmetric particle could constitute the cold dark matter,
indicated by the (red) diagonal hatching.

}
\label{fig:Susy}
\end{figure}

Most particle theorists believe that the Higgs boson will have to be
accompanied by other new particles, such as those predicted by
supersymmetry, which should also be accessible to the LHC, as seen in
Fig.~\ref{fig:Susy}~\cite{TDRs,Bench}.  The lightest supersymmetric
particle is a leading candidate~\cite{EHNOS} for the cold dark matter
thought by astrophysicists to infest the Universe.

LHC's lead-lead collisions will probe nuclear matter at temperatures and
pressures typical of those in the early Big Bang when the Universe was
less than about 10$^{-6}$~s old, when it is thought to have taken the form
of a quark-gluon plasma. Experiments at the LHC experiment will also
continue studies of matter-antimatter asymmetry in decays of particles
containing the bottom quark, probing whether they can be described by the
Standard Model, and seeking to cast light on cosmological baryogenesis. If
history is a reliable guide, the best-remembered discovery of the LHC will
probably be none of these!

The LHC experimental programme will consist of two major detectors,
ATLAS~\cite{ATLAS} and CMS~\cite{CMS}, that are designed to look for new
particles and other novel phenomena at high energies, another experiment
ALICE~\cite{ALICE} directed towards studies of heavy-ion collisions, and
the smaller LHCb experiment~\cite{LHCb} looking at matter-antimatter
asymmetry. New underground caverns have been dug to accommodate the ATLAS
and CMS experiments, whereas ALICE and LHC-b will be housed in caverns
used previously by LEP experiments, as seen in Fig.~\ref{fig:LHC}. Another
small experiment called TOTEM~\cite{TOTEM} will be attached to CMS to
measure the total and elastic proton-proton cross sections, and a
dedicated search for magnetic monopoles is also being
proposed.

For the first time, the LHC accelerator is being built as a true global
collaboration, with important contributions from many other laboratories
besides CERN. Outside CERN's member states, important components are being
provided by the United States, Russia, Japan, Canada and India. The
contracts for the 1200 main-ring dipole magnets have now been placed.
Pre-production models have already been delivered to CERN, and have met
the specifications for accelerating protons to 7~TeV. These magnets are
very challenging, since they must operate at 1.9~K and achieve fields of
9~Tesla. A test string of LHC magnets has been ramped successfully up to
the field required to reach this design energy.

Some delays were incurred in the civil engineering for the experimental
caverns and the tunnels used to transfer particles into the LHC ring,
which turned out to be quite complicated. For example, the access pit to
the CMS cavern had to be excavated through an underground stream, that had
to be frozen with liquid nitrogen before digging was possible. The cavern
itself consists of two large caves side-by-side. To support the rock
above, first the wall between these caves was excavated and filled with
concrete, and only subsequently could the caves themselves be dug out. The
ATLAS cavern is so large that its concrete roof has to be held up by steel
stays bolted into the rock above.

The civil engineering is now nearly complete, and no further delays are
anticipated from this source. The principal items still on the critical
path are procuring the superconducting cable for the main dipole magnets,
which has also incurred significant delays, and the cryolines in the LHC
tunnel. Because of the delays suffered so far, the scheduled completion of
the LHC ring has been pushed back to the end of 2006, with first
collisions in 2007.

\begin{figure}[h]
\begin{center}
\includegraphics[width=.7\textwidth]{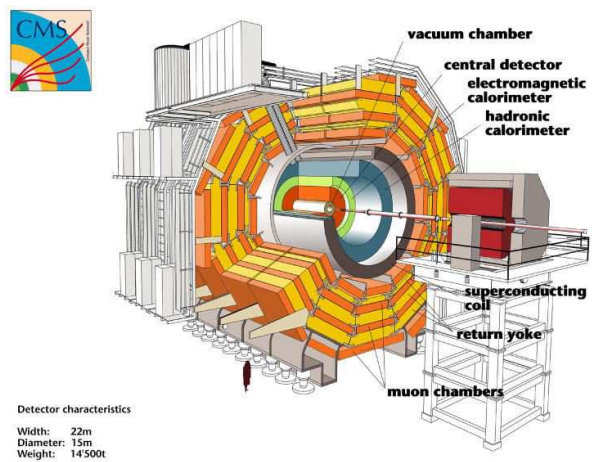}
\end{center}  
\caption[]{\it
Conceptual drawing of the CMS experiment, illustrating its `onion-skin' 
structure, with different layers designed to detect and measure different 
types of particles.
}
\label{fig:CMS}
\end{figure}

The major detectors resemble onions, as illustrated in 
Fig.~\ref{fig:CMS}, whose
concentric layers measure different types of particles produced in the
collisions, such as charged particles, photons, electrons,
strongly-interacting particles and muons. They include enough silicon
detectors to cover a football field, (in the case of CMS) tens of
thousands of lead tungstate crystals, and large superconducting magnets.
The ATLAS detector is so large that its cavern could accommodate CERN's
central administration building - and some wags have suggested that might
be a good place to put it. Large parts of these detectors have already
been constructed and delivered to CERN, ready for assembly and
installation in their respective caverns. The collaborations are on course
for having operational detectors ready to take useful data as soon as the
LHC starts colliding protons.

\begin{figure}[h]
\begin{center}
\includegraphics[width=.7\textwidth]{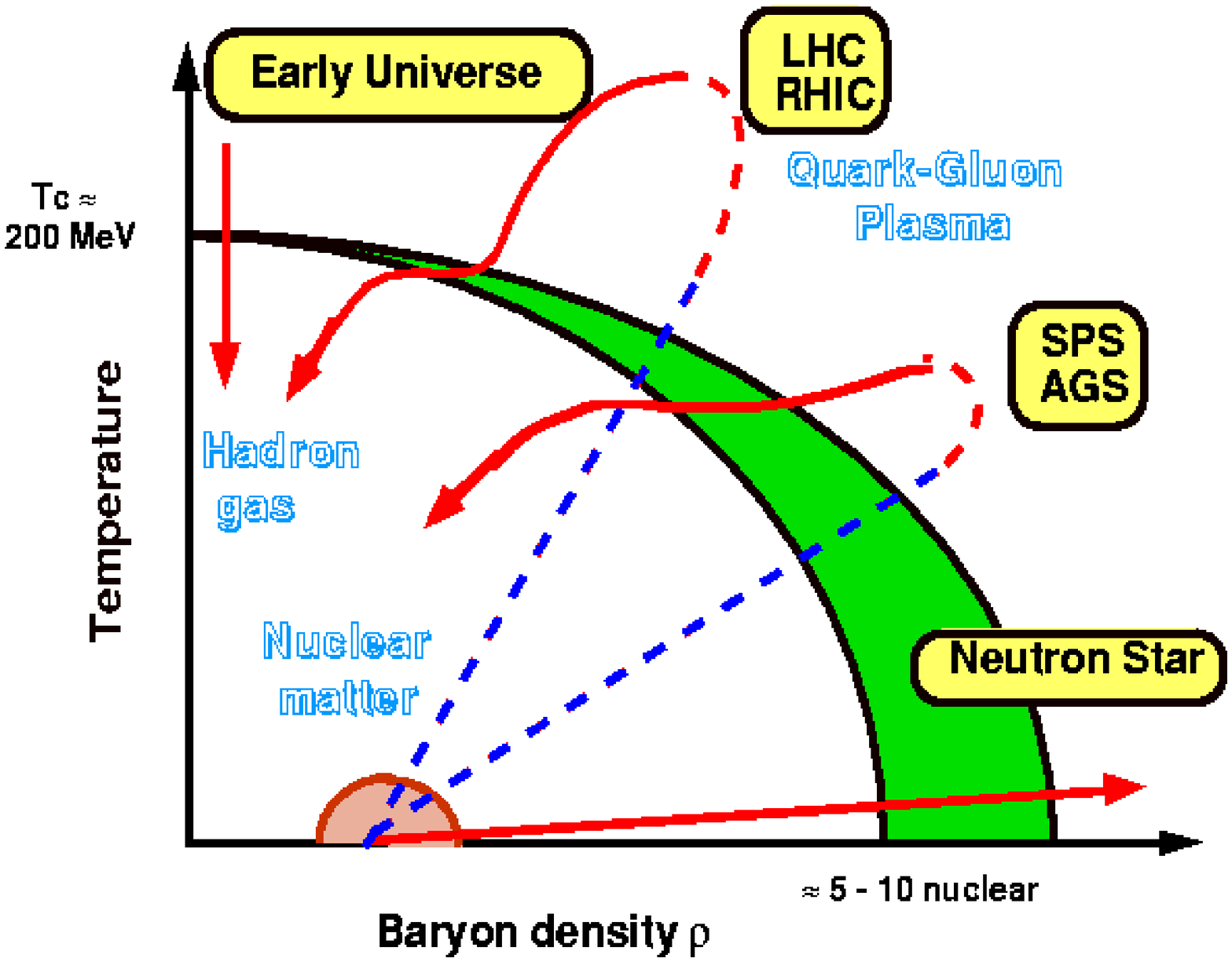}
\end{center}  
\caption[]{\it
Relativistic heavy-ion experiments such as ALICE at the LHC attempt to
produce a hot and dense state of matter, which is thought to have been a
quark-gluon plasma. The interiors of neutron stars are thought to be
relatively cool but with a large baryon density, SPS and Brookhaven AGS
experiments have probed higher temperatures, and the ALICE and Brookhaven
RHIC experiments will probe conditions closest to those in the first
microsecond of the Big Bang.
}
\label{fig:ALICE}
\end{figure}

The ALICE detector not only reuses an old LEP experimental cavern, it also
reuses the magnet used previously for the L3 experiment. Inside it also
has a large particle tracker and features specialized subdetectors for
photons, electrons and muons. The putative quark-gluon plasma is not
expected to have a single distinctive signature, or `smoking gun', but
rather is expected to be identified using a number of convergent
indicators, such as Hanbury-Brown-Twiss interferometry using particles
emitted from the last-scattering surface of the expanding fireball
produced by each `Little Bang', characteristic abundances of heavier
particles, and energetic photons emerging directly from inside the Little
Bang, reproducing conditions early in the Big Bang, as seen in
Fig.~\ref{fig:ALICE}.

Previous experiments at CERN and BNL colliding heavy-ion beams with fixed
targets provided matter under high pressures, but with
relatively low temperatures and high baryon densities, reminiscent of
neutron stars. More recent experiments with the RHIC heavy-ion collider at
BNL have pushed to higher temperatures~\cite{RHIC}, but the LHC will come
closest to reproducing conditions in the Big Bang, where the baryonic
chemical potential was negligible compared to the temperature.

\section{CERN's Global Network}

\begin{figure}[h]
\begin{center}
\includegraphics[width=.7\textwidth]{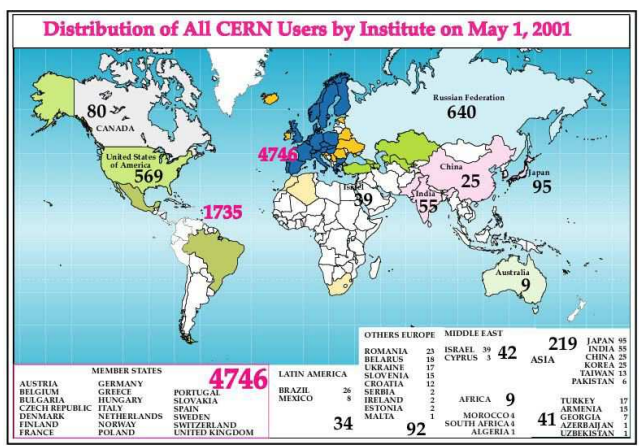}
\end{center}  
\caption[]{\it
High-energy physicists from around the world use CERN for their
experiments. Most come from CERN's European member states, but around
2000 scientists come from elsewhere.
}
\label{fig:world}
\end{figure}

The important contributions of non-European countries to the LHC
accelerator have already been underlined. The signicance of their
contributions to the LHC experiments is even greater: about 50\% of the
physicists and engineers in the experimental collaborations come from
institutions outside CERN's member states, and about 30\% of the total
value of the components of the detectors. Each of the major LHC detectors
has about 1800 participating physicists and engineers, and about fifty
coutries are represented officially. As seen in Fig.~\ref{fig:world}, the
total number of scientists registered as making scientific use of CERN's
facilities is between 6000 and 7000, with 4000 to 5000 coming from its
member states. The largest external contingents Russia and the United
States, around 600 each, followed by Japan, Canada, Israel, Brazil, China
and Korea. We collaborate with physicists from every continent except
Antarctica.

The age distribution of these scientific users is particularly
interesting: the mode is below 30, corresponding to the large numbers of
students and postdoctoral research associates that pass through CERN every
year. However, more than half of these students subsequently move out of
academic research, into industry, finance, etc..

\section{From the Web to the Grid}

The World-Wide Web was invented at CERN to enable all the far-flung
collaborators in its LEP experimental programme to share information.
Nobody at CERN anticipated the social phenomenon it would become. Now the
LHC project is confronting CERN with a staggering new computing challenge.

Each of the major LHC experiments will take data at a rate of about a
Petabyte ($10^{15}$ bytes) per second. This is equivalent to about a
billion people surfing the Web simultaneously, or everybody on the planet
making dozens of simultaneous mobile phone calls. The great majority of 
these
data are not interesting: perhaps only one collision in $10^{12}$ will
contain a Higgs boson, for example. Therefore, the detectors' data
acquisition systems are designed to discard all except one interesting
candidate in every $10^7$ collisions, approximately. This still leaves
several Petabytes of data to be recorded each year and analyzed, a task
that would require 100,000 or more of today's PCs, which CERN cannot
afford.

CERN plans to tackle this problem by deploying~\cite{LHCG} the Grid
computing technology, with which the user of a desk-top or portable
computer should be able to analyze data stored anywhere in the world,
using CPU power wherever it is available, just as we can switch a light on
without wondering where the electricity comes from. The Grid objective is
transparent user access to data, programs and computing power.
Applications of Grid technology to many other sciences, such as biology
(e.g., the human genome project), space science and environmental science
(e.g., earth observation) are also being developed, but the advent of the
LHC provides a definite time-frame over which particle physicists must get
a solution in place.

The amount of computing power required to simulate and analyze LHC data is
set to grow much faster than Moore's Law, the rate at which the computing
power of a single chip has grown historically. In the past, CERN has
solved this problem by establishing farms of commodity PCs, rather than
using monolithic supercomputers. The Grid project carries this
decentralization to its logical conclusion, linking together farms of
farms around the planet.

CERN has many collaborating partners in developing the Grid. One of the
largest European projects is DataGrid, funded by the European Union and
including ESA, PPARC in the United Kingdom, CNRS in France, INFN in Italy
and NIKHEF in the Netherlands as well as CERN. Other European projects
include CrossGrid, which involves new partners from Ireland and Central
Europe, there is the DataTag project to link up with the United States,
and parallel projects exist there and elsewhere. In the unique OpenLab
venture, we have also attracted industrial partners to provide hardware
for Grid development at CERN.

During Phase I of the LHC computing Grid project extending to 2005, the
plan is to write the basic software and middleware (system management
software), and to demonstrate its functioning with simulated data at a
level that is a significant fraction of the eventual LHC requirement.
Installation of the full LHC Grid is scheduled for the years 2006 to 2008.

\section{Options for CERN after the LHC}

\begin{figure}[h]
\begin{center}
\includegraphics[width=.7\textwidth]{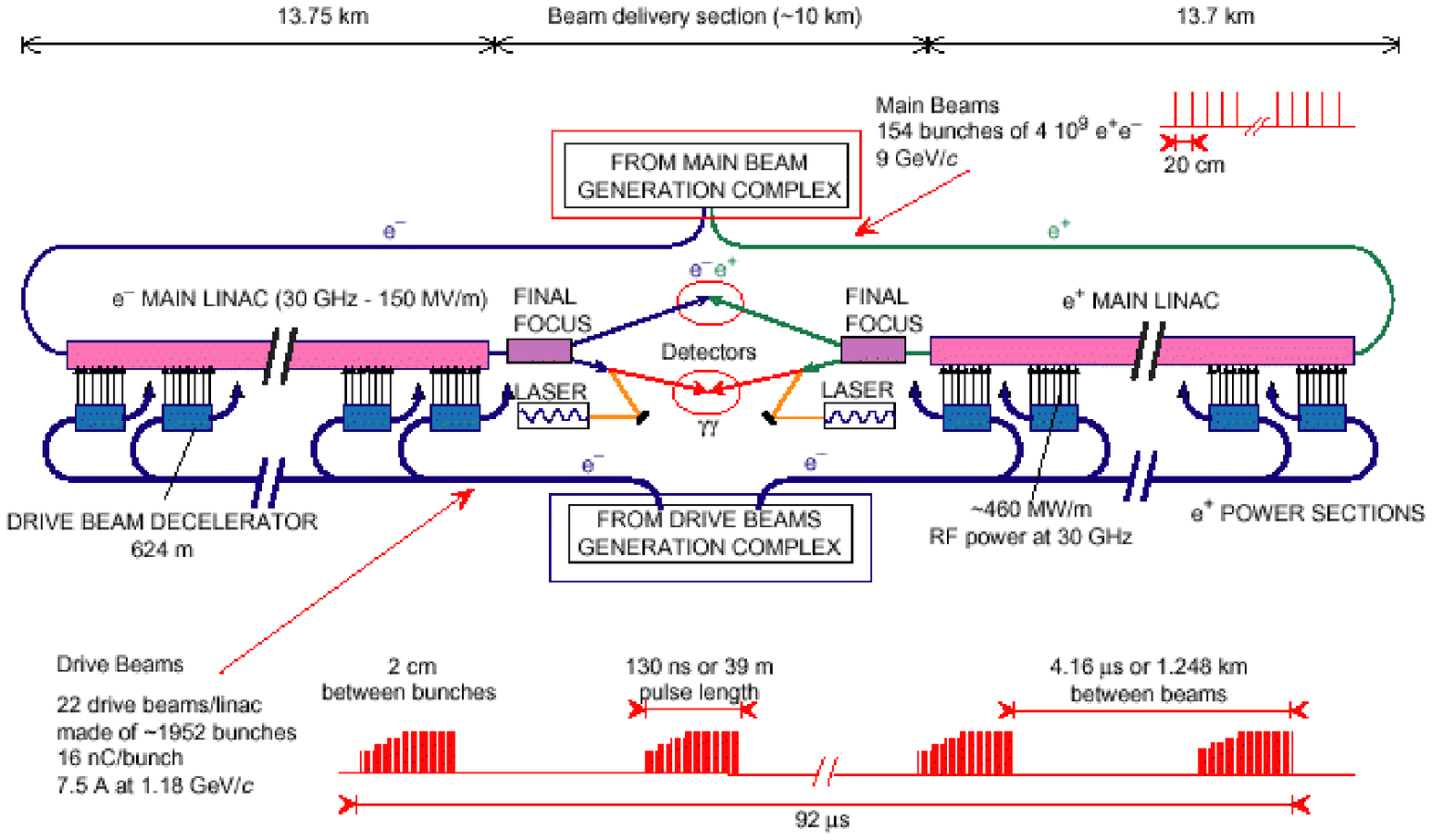}
\end{center}
\caption[]{\it
Conceptual design for a linear $e^+ e^-$ collider capable of reaching a
centre-of-mass energy of 3~TeV or more, called CLIC, which uses high-power
but low-energy drive beams to accelerate less-intense colliding beams to
high energies.
}
\label{fig:CLIC}
\end{figure}

Even though the LHC will only start taking data in 2007, and will continue
to provide exciting data for a decade or more, the preparation times
needed for new accelerator projects are so long that we at CERN have
already started thinking about possible new projects after
the LHC~\cite{EKR,DEG}. Several laboratories around the world have for 
some years already
been developing plans for linear electron-positron colliders capable of
centre-of-mass energies up to about 1~TeV, hoping to start
one of these projects in parallel with the operation of the LHC. One
of the options being considered at CERN is a possible next-generation
electron-positron collider called CLIC~\cite{CLIC}, capable of higher
centre-of-mass energies up to about 5~TeV: see Fig.~\ref{fig:CLIC}. If
supersymmetry exists, this would almost guarantee accurate measurements of
the properties of all the supersymmetric particles, enabling us, for
example, to calculate better the density of supersymmetric relics from the
Big Bang, and the rates at which they would scatter on conventional
matter.

In order to avoid a very long device, this energy objective would require
a very high accelerating gradient. CERN proposes to achieve this by an
innovative double-beam technique, in which an intense low-energy drive
beam is used to generate RF power that is then transferred to a
lower-intensity, higher-energy beam. This is the origin for the name CLIC,
for `Compact Linear Collider'. The double-beam principle has been
demonstrated in a couple of test facilities, and another now under
construction~\cite{CTF3} is intended to provide an engineering
demonstration of the CLIC concept. Another step towards demonstrating its
feasibility might be a Higgs factory using $\gamma \gamma$ collisions
generated by shining laser beams on $e^- e^-$ beams colliding at around 
centre-of-mass energies around 150~GeV~\cite{CLICHE}.

\begin{figure}[h]
\begin{center}
\includegraphics[width=.7\textwidth]{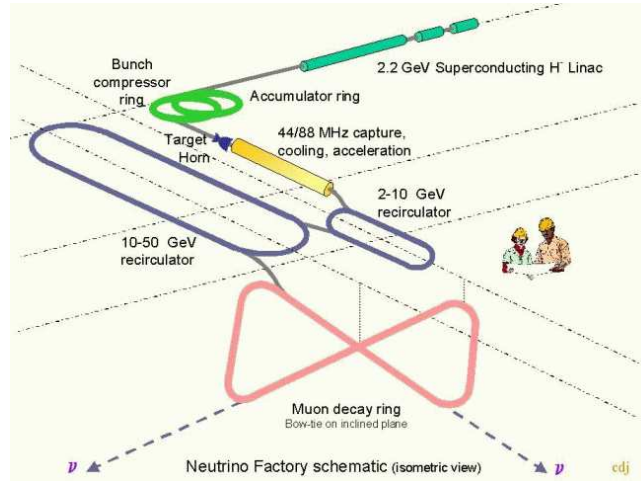}
\end{center}
\caption[]{\it
Conceptual layout for a neutrino factory, based on an intense
superconducting proton linac that produces many pions, whose decay muons
are captured, cooled in phase space and stored in a `bow-tie' ring. Their
subsequent decays send neutrinos with known energy spectra and flavours
to a combination of short- and long-baseline experiments.
}
\label{fig:nufact}
\end{figure}

Any new accelerator project is sure to require a global collaboration even
more widely spread than the LHC. However, other projects considered at
CERN have even more planetary dimensions. One possibility is a neutrino
factory~\cite{nufact}, designed to produce a controlled beam two or three
orders of magnitude more intense than current long-baseline projects,
using the decays of muons captured in a storage ring, as seen in
Fig.~\ref{fig:nufact}. One of the primary objectives of such a project
would be to measure matter-antimatter asymmetry in neutrino oscillations.
This might require sending a neutrino beam several thousand kilometres
from one continent to another, making the experiments truly global. Such
experiments might cast important light on one of the favoured scenarios
for baryogenesis~\cite{EHRS}, in which the decays of massive neutrinos
first provide a lepton asymmetry, that Standard Model interactions
subsequently convert into a baryon asymmetry. Following steps in this 
programme could include $\mu^+ \mu^-$ colliders with various centre-of-mass
energies, possibly including one of more Higgs factories and/or a 
high-energy frontier machine~\cite{nufact}.

Another future possibility for CERN that has attracted some interest is a
more active role in space experiments. As already mentioned, CERN
currently recognizes several space experiments, including AMS, CAPRICE,
GLAST and PAMELA. It has a similar interest in AUGER, which is looking for
ultra-high-energy cosmic rays, and EXPLORER, looking for gravitational
waves. The next steps in these directions may be EUSO~\cite{EUSO} or OWL,
monitoring from space cosmic-ray impacts in even larger volumes of the
atmosphere, and LISA. We at CERN certainly have legitimate scientific
interests in the physics of these experiments: once the problem of mass
has been sorted out, grand unification and gravity may be next on the
particle physicists' agenda, and space experiments such as these might
offer good ways to test them. But would CERN add significant value to such 
projects?

CERN will not take any irrevocable decision concerning its future before
we have at least some results from the LHC, providing clearer hints where
to head next. For now, CERN has its hands full constructing the LHC, with
all its technical, financial, organizational, computational and physical
challenges. This project provides us with us with exciting scientific
perspectives for the rest of this decade and (most of) the next, providing
us with plenty of time to develop one of the above ideas, or perhaps
another, into a longer-term perspective worthy of CERN's mission.

\end{document}